\documentclass[%
 reprint,
superscriptaddress,
 amsmath,amssymb,
 aps,
 pra,
]{revtex4-1}

\usepackage{graphicx}
\usepackage{dcolumn}
\usepackage{bm}
\usepackage[utf8]{inputenc}
\usepackage{amsmath,amssymb}
\usepackage{physics}
\usepackage[utf8]{inputenc}
\usepackage[english]{babel}
\usepackage{wrapfig}
\usepackage{float}
\usepackage{ftnxtra}
\usepackage{array}
\usepackage{enumitem}
\usepackage{lastpage}
\usepackage{titlesec}
\usepackage[bottom]{footmisc}
\usepackage{xcolor,colortbl}
\usepackage{fancyhdr}
\usepackage{braket}
\usepackage{csquotes}
\usepackage{caption}

\begin{document}

\preprint{APS/123-QED}

\title{Independent electrical control of two quantum dots coupled through a photonic-crystal waveguide}

\author{Xiao-Liu Chu}
 \altaffiliation[Present address:]{MRC London Institute of Medical Sciences, Du Cane road, London, W12 0NN, United Kingdom}
  \email{xchu@ic.ac.uk}
 \author{Camille Papon}
 \affiliation{%
 Center for Hybrid Quantum Networks (Hy-Q), Niels Bohr Institute,
University of Copenhagen, Blegdamsvej 17, DK-2100 Copenhagen, Denmark\\
}%
\author{Nikolai Bart}
\author{Andreas D. Wieck}
\author{Arne Ludwig}
\affiliation{Lehrstuhl für Angewandte Festkörperphysik, Ruhr-Universität Bochum, Universitätsstrasse 150, D-44780 Bochum, Germany}
\author{Leonardo Midolo}
\author{Nir Rotenberg}%
 \altaffiliation[Present address:]{Centre for Nanophotonics, Department of Physics, Engineering Physics \& Astronomy, Queen’s University, 64 Bader Lane, K7L 3N6 Kingston, Ontario, Canada }
 \author{Peter Lodahl}
\affiliation{%
 Center for Hybrid Quantum Networks (Hy-Q), Niels Bohr Institute,
University of Copenhagen, Blegdamsvej 17, DK-2100 Copenhagen, Denmark\\
}%


\date{\today}

\begin{abstract}
Efficient light-matter interaction at the single-photon level is of fundamental importance in emerging photonic quantum technology. A fundamental challenge is addressing multiple quantum emitters at once, as intrinsic inhomogeneities of solid-state platforms require individual tuning of each emitter. We present the realization of two semiconductor quantum dot emitters that are efficiently coupled to a photonic-crystal waveguide and individually controllable by applying a local electric Stark field. We present resonant transmission and fluorescence spectra in order to probe the coupling of the two emitters to the waveguide. We exploit the single-photon stream from one quantum dot to perform spectroscopy on the second quantum dot positioned  16$\mu$m away in the waveguide. Furthermore, power-dependent resonant transmission measurements reveals signatures of coherent coupling between the emitters. Our work provides a scalable route to realizing multi-emitter collective coupling, which has inherently been missing for solid-state deterministic photon emitters.

\end{abstract}

\maketitle

A challenge of modern quantum photonics is to scale up deterministic solid-state photon-emitter systems in order to couple multiple emitters. The challenge pertains to the inherent inhomogeneities of the systems, e.g., self-assembled quantum dots (QDs) suffer from significant morphological inhomogeneities during growth, resulting in spectral deviations of the individual emitters. Moreover spatial variations within photonic nanostructures imply different Purcell enhancement factors and therefore emitted photon wavepackets \cite{Lodahl:15}. Similar challenges are found for other  solid-state quantum emitters compatible with photonic nanostructures including organic molecules \cite{Faez:14} or single-defect centers \cite{Sipahigil:16}. Previous work on these platforms include the realization of high-quality single-photon sources \cite{Chu:17,Uppu:20}, entangled photon pairs \cite{Liu:19}, spin-photon interfaces \cite{Gao:12}, and coherent nonlinear optics  \cite{Hwang:09,Javadi:15,Liang:18}.

Recent progress has focused on scaling these systems up and coupling multiple emitters. Most previous work has considered quantum emitters in bulk samples, where high-quality photon-photon interference \cite{Zhai:22}  and near-field dipole-dipole coupling \cite{Trebbia:22} have been realized. In photonic nanostructures, super- and subradiant coupling through a  cavity/waveguide was recently observed \cite{Evans:18,Tiranov:23}, however independent tuning of each emitter as required for scalability was not yet achieved.
Various tuning mechanisms have been implemented based on strain \cite{Grim:19}, magnetic field \cite{Tiranov:23}, and electric field \cite{Papon:22}. The latter work realized independent tuning of QDs that were electrically isolated by etching shallow trenches into the device enabling quantum interference between two different waveguide single-photon sources.

\begin{figure*}
    \includegraphics[trim={0 0cm 0 0},clip, width=1\textwidth]{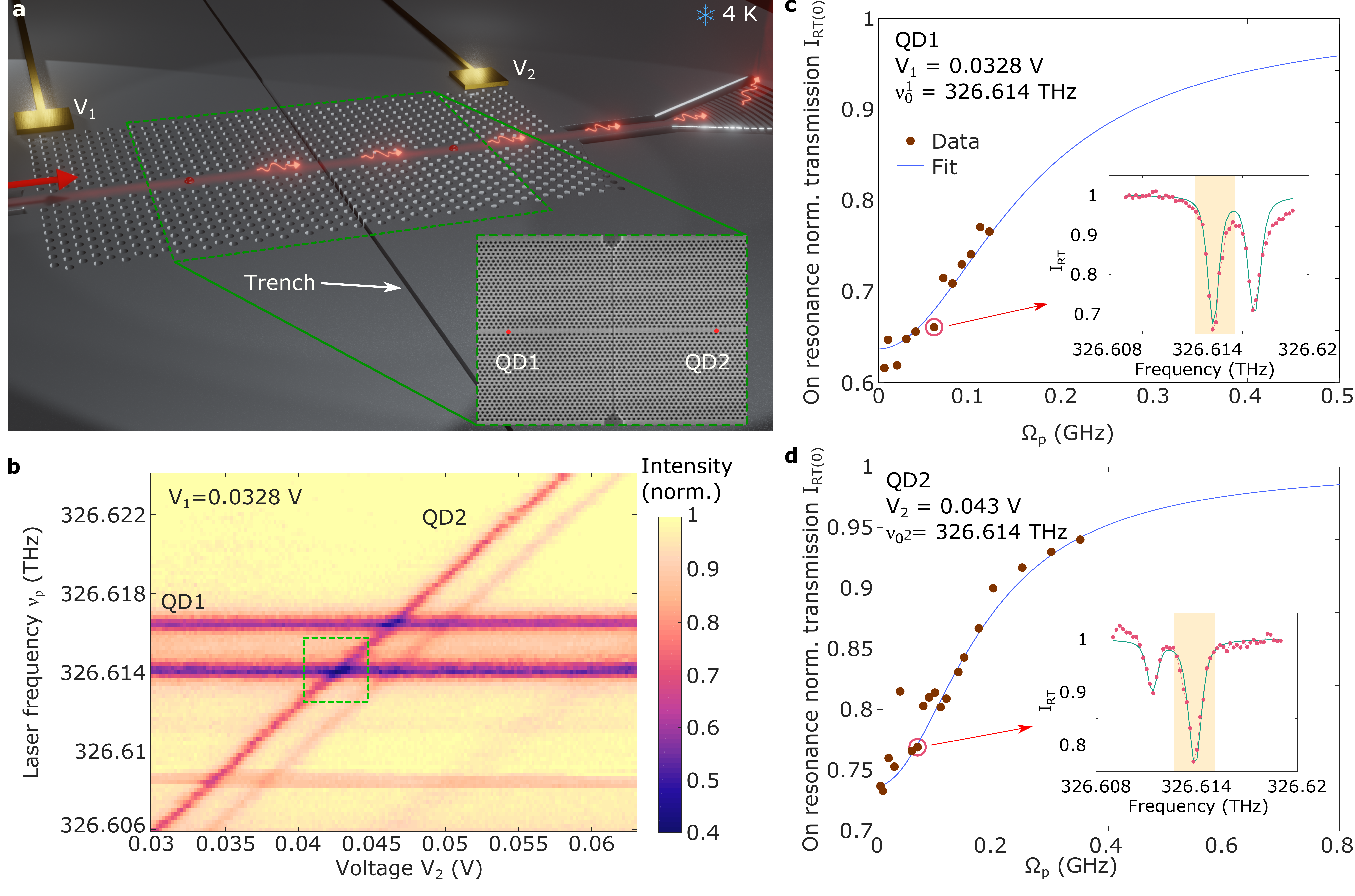}
    \caption{Independent electrical operation of two QDs in a PhCW. \textbf{a}, Epitaxially grown QDs are embedded within a PhCW on a p-i-n diode. The trench separates the waveguide, such that an independent gate voltage can be applied to each half, here illustrated as the metal contacts $\mathrm{V}_1$ and $\mathrm{V}_2$. Inset: An SEM image of the PhCW and corresponding 100 nm trench, with the rough positions of the two studied QDs marked ($15.7\pm 2 \mu m$ separation, see \cite{SM}). \textbf{b}, Independent control of the two QDs as demonstrated by keeping $V_1$ constant and measuring the transmission through the PhCW as $V_2$ is scanned. The corresponding resonant extinction dips corresponding to the two orthogonal dipoles (separated by about 2.4 GHz) of QD1 are independent of $V_2$, appearing as two horizontal lines on this map, while those of QD2 vary linearly with the gate voltage. From this map, an overlap of two QD resonances near 326.614 THz is identified (green dashed region). \textbf{c, d}, Measured power-dependent resonant extinction for QD1 and QD2, respectively, when the other QD is far-detuned. The corresponding theoretical fits (see \cite{SM}) are used to determine the individual QD parameters. The insets show exemplary RT scans, with $\mathrm{I_{RT}}$} defined as the extinction dip depth and the transitions that are tuned into resonance are highlighted by the shaded regions.
  \label{Fig:setup}
\end{figure*}

Here, we extend the multi-emitter work and realize independent tuning of QDs efficiently coupled to the same  photonic crystal waveguide (PhCW), cf. schematic illustration in Fig.~\ref{Fig:setup}a. PhCWs are excellent quantum photonic platforms, enabling near-unity light-matter coupling \cite{Arcari:14} by suppressing emission into free-space \cite{Mango:07} over a broad wavelength range, and enabling near-transform-limited optical transitions when the emitters are embedded in a p-i-n diode \cite{Pedersen:20}. We realize a PhCW device that is divided into two halves by etching a 100 nm wide and $~$50 nm shallow trench into the p-doped layer, whereby two halves are electrically isolated and therefore individually Stark tunable \cite{Bennett:10,Kirsanske:16}, as controlled by gate voltages $V_1$ and $V_2$, respectively. The trench is designed to have a minimal effect on the optical waveguide mode, which is consistent with the fact that no optical scattering is observed when imaging the trench region of the waveguide. The demonstrated approach could readily be extended to control additional QDs and therefore provides a route of scaling up the platform. We present  resonant fluorescence and resonant transmission data while individually tuning each QD resonance frequency and implementing selective optical excitation.

We start by demonstrating independent electrical tuning of QDs within the same PhCW by measuring the emitter resonance frequency as a function of an externally applied gate voltage. The QDs are optically excited with a tunable continuous wave (cw) laser: i) either directly through the waveguide in a resonant transmission (RT) experiment, in which case both emitters are simultaneously excited via the guided mode  or ii) each QD is selectively excited from free-space in a resonance fluorescence (RF) experiment. In both experimental configurations, shallow-etched gratings \cite{Zhou:18} are used to efficiently couple light out from the waveguide mode, as sketched in Fig.~\ref{Fig:setup}a. 

We begin by acquiring an RT spectrum to first identify the QDs that efficiently couple to the waveguide \cite{Thyrrestrup:18}. We show an exemplary transmission map in Fig.~\ref{Fig:setup}b, where scattering from the emitters results in an extinction of the detected signal as these photons are predominantly reflected  \cite{Turschmann:19}. For this measurement, the gate voltage on the left section of the PhCW $\mathrm{V}_1$ is held constant, while $\mathrm{V}_2$ on the right-hand side is varied. Indeed, we observe that by varying $V_2$, only the transition resonances associated with QD2 tune, while those of QD1 remain constant, i.e. they appear as horizontal lines in Fig.~\ref{Fig:setup}b. That is, the device enables individual electrical tuning of the QDs, allowing to bring individual emitters into mutual resonance, as highlighted by the green dashed curve. Note that these two QDs are separated by $15.7\pm 2 \mu m$ (see the \cite{SM}), corresponding to 22 $\pm$ 2 times the wavelengths in the PhCW for the estimated group velocity in the waveguide of 2 ($\pm 0.6$)$\times 10^7$.
\begin{figure*}
    \includegraphics[trim={0 0cm 0 0},clip, width=1\textwidth]{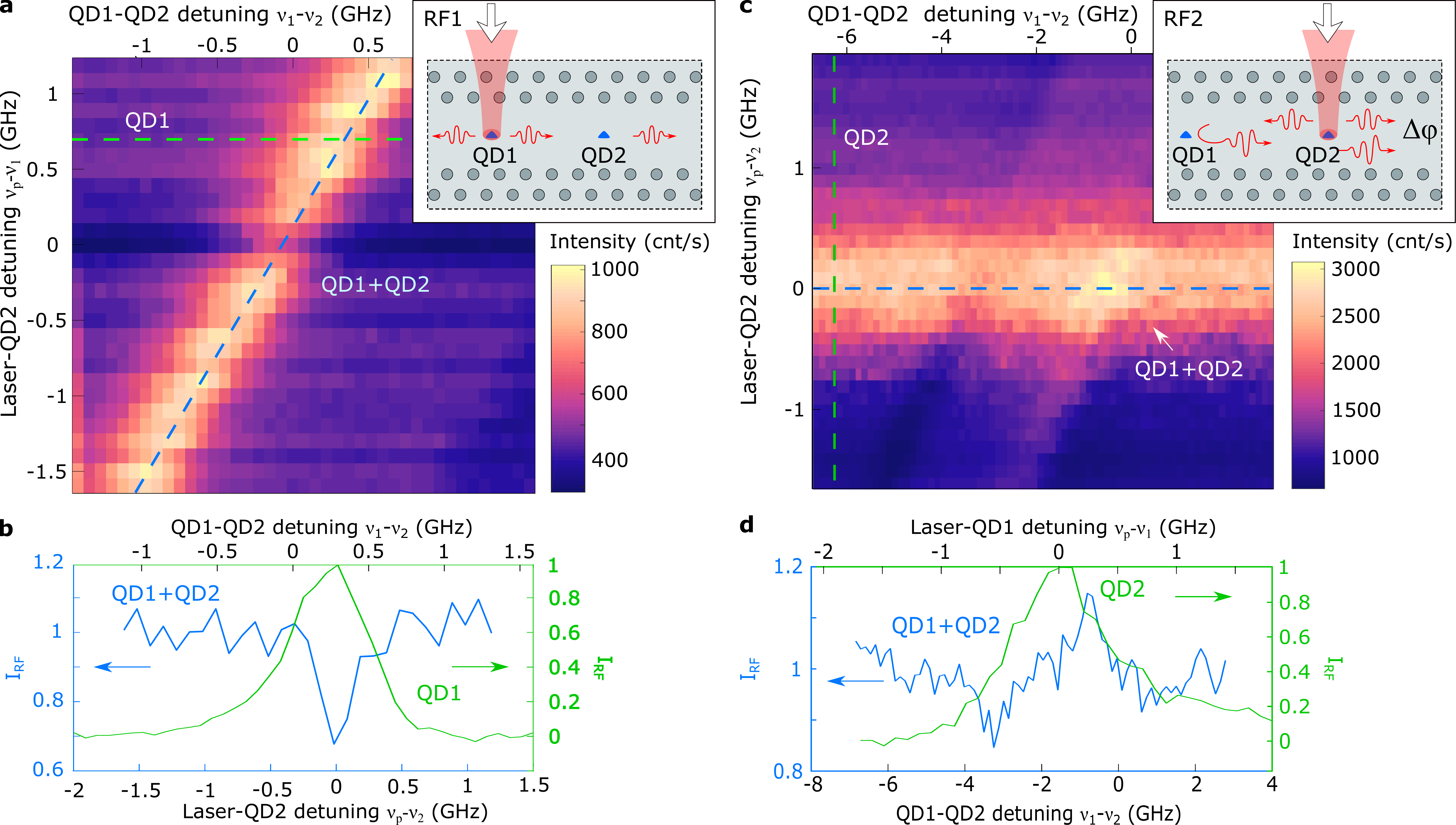}
    \caption{Independent optical excitation of each QD in a RF configuration, where the photons emitted from one QD travel through the waveguide to interact with the second emitter. \textbf{a} In RF1, the cw laser is scanned through the QD1 resonance, while voltage-tuning $(\mathrm{V}_1)$ across the QD2 frequency and simultaneously keeping the QD2 resonance constant at $\nu_2=$ 326.614 THz. The emitted single photons from QD1 scatter off QD2 and the resultant transmission spectrum reveals the coherent extinction of the QD2 resonance. \textbf{b} Cross-sections taken along the blue and green dashed curves in \textbf{a}, showing the RF spectrum of QD1 when QD2 is far detuned and the response of the two-QD system when the QDs are on resonance, respectively. \textbf{c} In RF2, the detected spectrum results from the interference of photons that are initially emitted from QD2 towards the right interfered with photons emitted to the left and subsequently reflected from QD1, see illustration in the inset. This signal depends on the relative phase $\Delta \varphi$ gained by the scattered photons (see \cite{SM}) \textbf{d} Same as in \textbf{a}, except that the green curve shows the RF spectrum of QD2.}
  \label{Fig:rfscan1}
\end{figure*}

Two resonances, associated with the in-plane orthogonal transition dipoles of InAs QDs \cite{Lodahl:15}, appear as parallel extinction lines for each emitter. For each QD, we select one transition (dashed green region in Fig.~\ref{Fig:setup}b), detune the other QD and measure RT for different excitation strengths. Figures~\ref{Fig:setup}c and d present the power-dependent normalized transmission $I_{\mathrm{RT}\left(0\right)}$ (see inset) for the two QDs, where the subscript $0$ indicates that the emitter detuning is zero, here plotted against the Rabi frequency $\Omega_p$ (see \cite{SM} for details). In both cases, the extinction decreases as the excitation power increases, as expected from theory (solid curves; see \cite{SM} for more details on the model) \cite{Novotny:06}. We extract that both QDs are efficiently coupled to the PhCWs, with coupling coefficients $\beta_1 = 0.85 \pm 0.05$ and $\beta_2 = 0.78 \pm 0.09$, while residual slow spectral diffusion limits their coupling, see \cite{SM} for full QD parameters. These results therefore demonstrate that the addition of a shallow trench allows us to electronically address QDs independently within the same waveguide mode.

To address each QD individually, we excite the QDs from free-space as shown in the insets of Fig.~\ref{Fig:rfscan1}a and c. The excited QD1 emits a stream of single photons into the waveguide, which travels through the waveguide and scatters coherently off the second QD2. The signal, $I_{\mathrm{RF}}$, is measured at the output port on the right after QD2. The system is operated at a relatively high excitation intensity (Rabi frequency $\Omega_p = 0.16$ GHz, or 3.3 photons per lifetime) in order to combat the effect of residual spectral diffusion, i.e. power broadening is also observed. We start by considering QD1 as the photon source. The single photons are emitted by QD1 resonantly excited at the laser frequency $\nu_{\mathrm{p}}$ and travel in both directions in the waveguide. We record photons traveling towards the right after they interact with QD2 by measuring $I_{\mathrm{RF}}$. Figure~\ref{Fig:rfscan1}a shows a map of the signal as a function of the laser-QD1 detuning ($\nu_{\mathrm{p}} - \nu_1$) and QD1-QD2 detuning ($\nu_1 - \nu_2$), of which the latter is voltage-controlled. When the two QDs are detuned, we measure $I_{\mathrm{RF}} = 1$ kcnts/s and observe a linewidth of $0.49 \pm 0.03$ GHz in Fig.~\ref{Fig:rfscan1}b (green curve). When QD2 is tuned into resonance with QD1, it coherently reflects photons, leading to an extinction of the transmission. This effect can be seen in the cross-section data along the dashed blue line in Fig.~\ref{Fig:rfscan1}a, which is plotted in blue in Fig.~\ref{Fig:rfscan1}b. In this configuration, the recorded signal is independent of the phase lag associated with propagation between the two QDs, since this constitutes a global phase shift not affecting the intensity measurements (See \cite{SM} for more information).

A second and more complex scenario arises when QD2 acts as the single-photon source; here the measured signal is comprised of both the photons that are initially emitted to the right by QD2 and those reflected by QD1, as illustrated in the inset to Fig.~\ref{Fig:rfscan1}c. Given the coherence of the photons scattered by QD1, the relative phase gain on the round trip between the emitters, $2\Delta \varphi$, as determined by the separation of the QDs, directly influences the $\mathrm{I}_{\mathrm{RF}}$ signal, as seen in Fig.~\ref{Fig:rfscan1}c and d. A complex Fano-like lineshape in $\mathrm{I}_{\mathrm{RF}}$ is observed when both QDs are tuned into resonance (see blue curve in Fig.~\ref{Fig:rfscan1}d), and the data are consistent with an overall phase separation of $\Delta \varphi \simeq 200\pi/180$, which sensitively determines the spectral shape of the signal, see \cite{SM} for the analysis. This phase separation determines the dispersive/dissipative character of collective interaction between coherently coupled emitters \cite{Chu:22}. The present experiment demonstrates coherent scattering on a single quantum emitter of single photons emitted by another emitter. 

\begin{figure}[!ht]
  \begin{center}
    \includegraphics[trim={0 0cm 0 0},clip, width=0.5\textwidth]{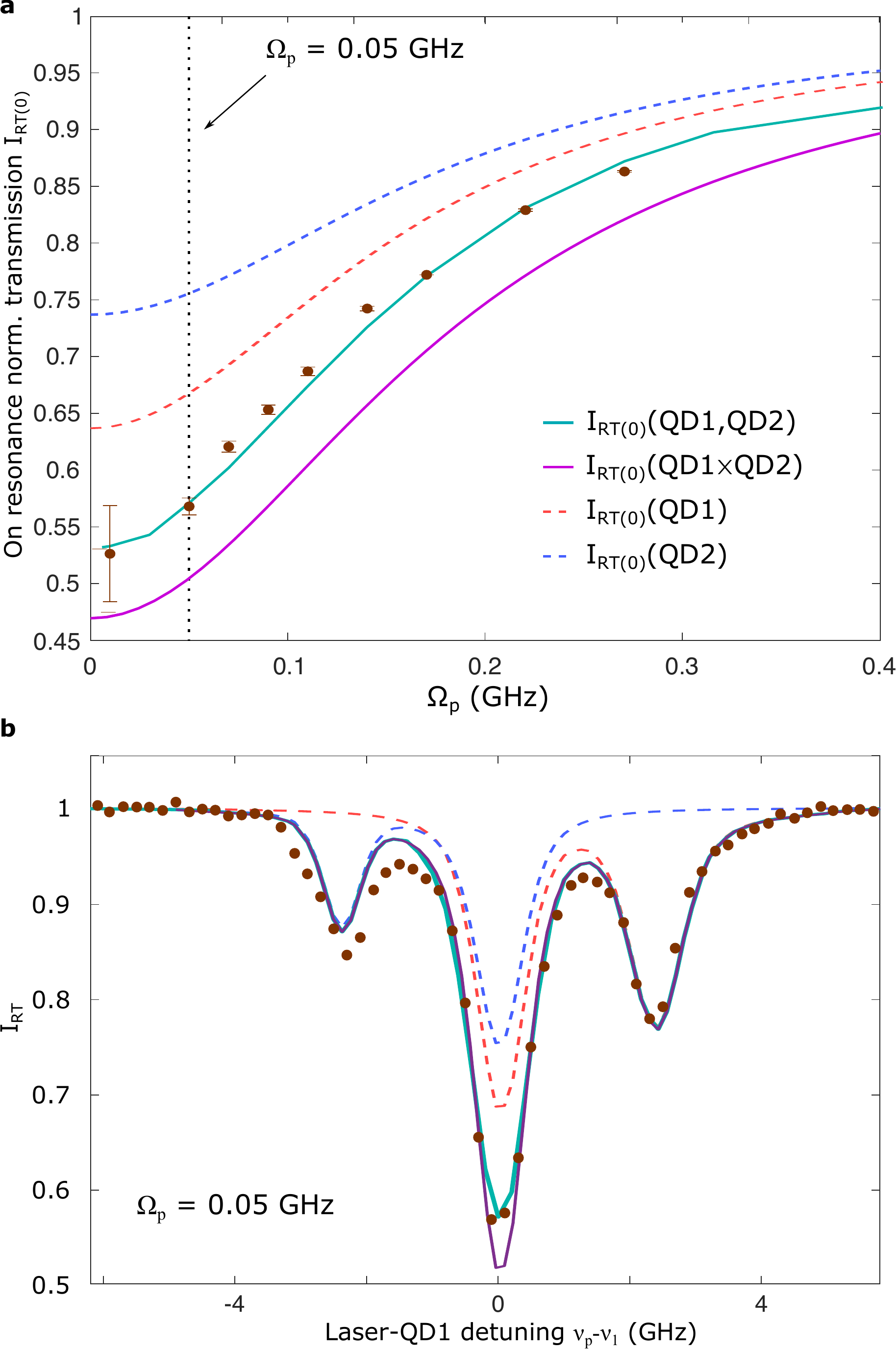}
  \end{center}
    \caption{Simultaneous optical excitation of both QDs through a single photonic waveguide. \textbf{a} Power-dependent transmission through the waveguide when the two QDs are both on resonance. Overlayed on the measurements (circular markers, errors due to detector dark counts) are models of $I_{\mathrm{RT}}$ for QD1 and QD2 alone (dashed red and blue curves, respectively) as well as their joint response assuming they either are independent or coherently coupled (solid purple and teal curves, respectively). Note that all parameters used in the models are determined from previous experiments. \textbf{b} An exemplary RT spectra of the two QDs on resonance overlayed with theoretical curves for the coupled and uncoupled response. The individual QD frequency scans are also added for comparison.}
  \label{Fig:rtscan2}
\end{figure}

Finally, we return to the original RT configuration as shown in Fig.~\ref{Fig:setup}a, tune both QDs into resonance and study the saturation behavior of the device. The joint resonance is probed at different excitation powers and the resultant photon flux-dependent transmission is plotted in Fig.~\ref{Fig:rtscan2}a. As is the case for a single emitter \cite{Hwang:09,Javadi:15,jeannic:21}, the extinction decreases non-linearly with increasing excitation flux, as the emitter saturates. For the two-QD system, a peak extinction of about 0.5 is observed, which is stronger than the single-emitter response (dashed curves in Fig.~\ref{Fig:rtscan2}a).

We overlay the modelled transmission of the two-QD system in the case where the emitters are uncoupled (purple curve) and coherently coupled (cyan curve) in Fig.~\ref{Fig:rtscan2}a. Here, the system parameters extracted from earlier experiments (c.f. Table I in \cite{SM}) and $\Delta\varphi \simeq 200\pi/180$ are used without any additional adjustable parameters and remarkably good agreement between theory and experiment is observed (see \cite{SM} for further details). We observe that the data best agree with the predictions for a coupled system, suggesting that the QDs couple via the PhCW mode. This is reinforced by the individual RT spectra, an example of which is shown in Fig.~\ref{Fig:rtscan2}b for the case of $\Omega_p = 0.05$ GHz. Here, the measured spectrum (circular markers) is overlayed with predictions for both the coupled and uncoupled system (showing the response of the individual QDs as dashed curves), observing that the uncoupled emitter theory  overestimates the total extinction. 

We have presented a quantum photonic system consisting of a PhCW that has been trenched to create two separate diode regions, each of which contains an efficiently coupled, high-quality QD. We are consequently able to address each QD separately, both with a local electric Stark field and with multiple optical excitation pathways. The system constitutes a versatile platform for multi-emitter experiments and technologies. Using this platform, we observe coherent scattering from both QDs, both when one is excited from free-space, and when both simultaneously interact with photons travelling through the guided mode. Our system therefore opens a window to the rich physics of cooperative quantum light-matter interactions in multiple-emitter systems.

This integrated platform can be readily scaled, for example by trenching the PhCW into additional sections, hence facilitating independent electric control of multiple emitters within the same waveguide. The approach paves a scalable route towards realizing many-emitter coherent and deterministic radiative coupling enabling creating scalable sub- or super-radiant collective quantum states \cite{Asenjogarcia:17b,Albrecht:19}, accessing decoherence free sub-spaces for quantum computation \cite{Paulisch:16} or to realize complex photonic cluster states for quantum communications \cite{Economou:10}.

\begin{acknowledgments}
The authors acknowledge financial support from Danmarks Grundforskningsfond (DNRF 139, Hy-Q Center for Hybrid Quantum Networks) and the EU's Horizon 2020 research and innovation programme (grant No. 824140, TOCHA, H2020-FETPROACT-01-2018). NR acknowledges funding from the Canadian Foundation for Innovation (CFI) and the Natural Sciences and Engineering Research Council of Canada (NSERC). 
\end{acknowledgments}


\end{document}